\documentclass[letterpaper,twocolumn,10pt]{article}
\usepackage{usenix,epsfig,endnotes}
\usepackage{graphicx}
\usepackage{xspace}
\usepackage{tikz}
\usepackage{float}
\usepackage{listings}
\usepackage{geometry}
\usepackage{xcolor}
\usepackage{parskip}
\usepackage{titlesec}
\usepackage{setspace}
\usepackage{hyperref}

\lstdefinestyle{customc}{
  abovecaptionskip=1\baselineskip,
  breaklines=true,
  frame=L,
  xleftmargin=\parindent,
  language=C,
  showstringspaces=false,
  basicstyle=\footnotesize\ttfamily,
  keywordstyle=\bfseries\color{green!40!black},
  commentstyle=\itshape\color{purple!40!black},
  identifierstyle=\color{black},
  stringstyle=\color{red},
  captionpos=b,
}

\titlespacing*{\subsubsection}
{0pt}{1.5ex plus 1ex minus .2ex}{1.5ex plus .2ex}
\titlespacing*{\section}{0pt}{1.5ex plus 1ex minus .2ex}{1.3ex plus .2ex}
\titlespacing*{\subsection}{0pt}{1.5ex plus 1ex minus .2ex}{1.3ex plus .2ex}

\newcommand\PN[1]{\paragraph{#1}}

\def\Snospace~{\S{}}

\newcommand*\BC[1]{%
\begin{tikzpicture}[baseline=(C.base)]
\node[draw,circle,fill=black,inner sep=0.1pt](C) {\textcolor{white}{\scriptsize #1}};
\end{tikzpicture}}

\newcommand\cc[1]{\texttt{#1}}
\newcommand\ie{\textit{i.e.}}
\newcommand\eg{\textit{e.g.}}
\newcommand\etal{\textit{et al.}}
\newcommand\sys{eBPF\xspace}

\title{\Large \bf The eBPF Runtime in the Linux Kernel}

\author{
{Bolaji Gbadamosi}\\
Karlstad University
\and
{Luigi Leonardi}\\
University of Pisa
\and
{Tobias Pulls}\\
Karlstad University
\and
{Toke Høiland-Jørgensen}\\
Red Hat
\and
{Simone Ferlin-Reiter}\\
Red Hat
\and
{Simo Sorce}\\
Red Hat
\and
{Anna Brunström}\\
Karlstad University
}
\date{August 2024}

\begin{document}

\maketitle

\begin{abstract}

Extended Berkeley Packet Filter (\sys) is a runtime that enables users to load programs into the operating system (OS) kernel, like Linux or Windows, and execute them safely and efficiently at designated kernel hooks. Each program passes through a verifier that reasons about the safety guarantees for execution. Hosting a safe virtual machine runtime within the kernel makes it dynamically programmable. Unlike the popular approach of bypassing or completely replacing the kernel, \sys gives users the flexibility to modify the kernel on the fly, rapidly experiment and iterate, and deploy solutions to achieve their workload-specific needs, while working in concert with the kernel.

In this paper, we present the first comprehensive description of the design and implementation of the \sys runtime in the Linux kernel. We argue that \sys today provides a mature and safe programming environment for the kernel. It has seen wide adoption since its inception and is increasingly being used not just to extend, but program entire components of the kernel, while preserving its runtime integrity. We outline the compelling advantages it offers for real-world production usage, and illustrate current use cases. Finally, we identify its key challenges, and discuss possible future directions.

\end{abstract}

\section{Introduction}
\label{s:intro}

Contemporary monolithic operating systems like Linux are designed to be
general-purpose, and cater to a wide variety of users. They define the necessary
abstractions to multiplex and share hardware resources safely and efficiently.
As such, their design choices play a key role in influencing an application's
performance, scalability, and security.

Evidently, the monolithic design choice of the Linux kernel does come with a
cost such as increased complexity and maintenance challenges due to tightly coupled components, higher security risks from a larger attack surface \cite{phdthesis}, reduced scalability and flexibility in resource-constrained environments.

As a result of these challenges, the approach of bypassing or replacing the OS kernel has gained
momentum. Kernel bypass solutions \cite{dpdk, spdk, demikernel:sosp21} and library OS~\cite{shenago} allow specializing the entire OS stack for a specific
workload, giving significant performance improvements. However, these solutions
may inhibit resource multiplexing by taking complete control of the hardware. They
may also require application logic to be rewritten, and give up traditional benefits
of an OS like the security and isolation model.

Users managing a large fleet of machines care deeply about extracting maximum
performance and utilization out of their hardware, while running their
production workloads on a battle-tested OS kernel like Linux to
reduce their maintenance burden. Ideally, they wish to have comparable
performance to prior approaches without abandoning their well understood tooling
for performance monitoring, administration, and management.

To this end, another direction is to explore tailoring the kernel mechanisms and
policies for specific workloads, which can yield drastic improvements. However, this is far from trivial on Linux. It involves changing
the kernel code, and deploying workload specific kernel changes to a large set
of machines, which is impractical at scale, due to the huge variety of target
applications, the complexities of frequent kernel redeployments.
There has been various research projects for safer ways to extend the kernel. For instance, VINO (Virtual Integrated Network Operating system)  \cite{small1994vino} and SPIN \cite{bershad1995extensibility} allow users to customize their kernels through user-defined functions, SPIN is written in Modula-3 language to make sure that dynamically loaded modules are well-secured and efficient. On the other hand, VINO primarily focus on how extensions can be put together safely via fault isolation techniques within them. However, the TockOS (Tock Operating System) \cite{culic2022low} is a more recent initiative that uses Rust to improve security and dependability in its microkernel design for embedded systems.
 Unlike these systems which often require new operating environments or adjustments, eBPF (Extended Berkeley Packet Filter) allows dynamic programming capabilities directly into monolithic kernels like Linux and Windows. As opposed to programs such as VINO, SPIN and TockOS, this integration enables eBPF develop secure enhancements without requiring a new kernel structure. Introduced in the kernel 3.18 (released on December, 2014) \textbf{\sys} functions as a \textbf{safe} programmable virtual
machine hosted on top of a \textbf{performant} in-kernel runtime. It allows
users to write programs and load them into the kernel, and attach them at
designated hooks to begin execution.
To ensure safety, every program is statically checked by a \textbf{verifier}
when being loaded. To ensure performance, all programs are \textbf{Just-in-time (JIT) compiled}
to native machine instructions.

In essence, \sys makes the Linux kernel \emph{dynamically programmable at
runtime}, while ensuring its runtime integrity remains intact. However, thus far, since its initial release until version 6.7 (released on January, 2024), which this paper is based on, there exists no complete description of the design and implementation of
\sys within the Linux kernel. Hence, through this paper, we make the following contributions:

\begin{itemize}
    \item A comprehensive description of the design and implementation of the
	  \sys runtime in the Linux kernel up until version 6.7.
    \item An exhaustive characterisation of \sys's safety properties (\autoref{s:safety}).
    \item Identification of limitations and key challenges concerning \sys's current design (\autoref{s:challenges}).
\end{itemize}

The rest of this paper is organized as follows. \autoref{s:background} presents background, and \autoref{s:overview} is about the general overview of \sys program. In \autoref{s:workflow} we illustrated the high-level programming and execution model of \sys. We discuss the four major passes involved during the verification of an \sys program in \autoref{ss:cfg},\autoref{ss:symbex},\autoref{ss:opt},\autoref{ss:jit}. We then present some use cases in \autoref{s:usecases}, challenges in \autoref{s:challenges} and conclude the paper in \autoref{s:conclusion}.

For brevity, the rest of the paper uses `\sys' to refer to the Linux runtime, as is common in the Linux kernel community.

\section{Background and Design Principles}
\label{s:background}

In recent years, Linux kernel customization has become a prevalent need across different sectors, with regard to performance, security, 
and observability goals. The last two decades have seen Linux customization evolve 
from being only a peripheral consideration to a critical necessity. in order to 
align  the kernel behavior with various operational requirement.

\subsection{Challenges with Kernel Customization}
\label{ss:challenges}
Developers often encounter significant hurdles when they need to make direct changes to the Linux kernel. Below are some of the key challenges.
\PN{Changing the Kernel} Linux provides a huge number of configuration knobs
both during compilation and runtime~\cite{kconfig,sysctl}. However, these
knobs do not fundamentally address performance bottlenecks or provide insight
into the kernel's behavior. They are appropriate for tuning parameters of
existing kernel policies, but not expressive enough for encoding new ones.

In such cases when customizing the kernel is warranted, developers would need to
make changes to the kernel code, and/or write kernel modules. However, this
makes testing and debugging changes much harder. Making involved changes to the
kernel requires deep familiarity with its complex codebase, incurring a huge
maintenance burden. Unless such changes are accepted by the kernel community,
they need to be forward ported on kernel upgrades due to unstable kernel APIs.
Any uncaught bugs in the code could easily lead to \emph{system crashes} and take
down production servers, which directly translates to downtime and lost revenue.

\PN{Deploying Kernel Changes} Deploying a change to the kernel, however small,
across a fleet of servers is a lengthy process~\cite{metadeploy}. Replacing the
running kernel with a new one involves disruption of workloads hosted on the
machine, as they must be stopped. The machine must then boot into the new
kernel, and reinitialize all services again. This cycle is particularly harmful
for workloads which incur cold starts and have a significant ramp-up time. After
all these steps, an extensive testing phase needs to be carried out to test and
qualify these changes, while fixing any detected regressions. This process has
to be repeated for every server which receives a kernel update, making the
process expensive and taking the total time to rollout a new kernel to the fleet
in the order of weeks or months.

\subsection{Alternative Approaches and Limitations}
As mentioned in \autoref{ss:challenges} developers face complexities when making changes to the kernel. We discuss alternative methods that promise performance benefits but also pose limitations and maintenance challenges.
\PN{Completely Bypassing or Replacing the Kernel.}
Although kernel bypass frameworks and
library operating systems specialized for a
particular workload are appealing in terms of performance, they come with their
own downsides. Kernel bypass solutions require taking complete control of the
hardware, and waste CPU cycles in busy polling. Library operating systems are
tailored for specific workloads, and require code modifications to adapt to
different requirements. In addition, both of these approaches make it difficult
or impossible for two workloads to coexist and share hardware resources, which
is not acceptable for some users. Finally, they require rewriting application
logic, which is a significant maintenance burden.

eBPF achieves these requirements by allowing efficient customization of the
Linux kernel safely and dynamically at runtime, reducing the risk of
introducing kernel bugs when making changes to the kernel, and accelerating
development and deployment velocity.
Prior to \sys's introduction, the kernel had multiple domain-specific
register-based virtual machines serving dedicated use cases in the networking
subsystem~\cite{04Bertone2018},
but none of them were designed to be general-purpose\footnote{Unlike the widely accepted misconception, \sys's design was \emph{not}
	 influenced by Classic BPF~\cite{cBPF,mccanne1993bsd},
	 and the name was only chosen for familiarity~\cite{bpfstory}.}.

\subsection{Design Principles}
We now outline the key design principles behind \sys's continued development and
evolution.

\PN{Safe and Dynamic OS Customization}
Hosting a safe and efficient virtual machine runtime within the kernel allows
customizing the kernel dynamically, achieving performance similar to existing
in-kernel code, while ensuring that the kernel's integrity is not undermined
under any circumstances. Programming against a safe virtual machine with a
constrained environment allows lowering the barrier to making modifications to
the kernel. The guarantee of safety instills greater confidence when such
programs are deployed to the kernel, as opposed to direct changes to the kernel
or kernel modules reducing the risk of introducing kernel bugs that could lead to service outages.

\PN{Rapid Deployment and Upgrades}
Changing the kernel and deploying it across a large fleet of servers is
significantly expensive, since the reboot of a machine incurs the cost of
killing and reinitializing the workload. Loading programs into the kernel
dynamically at runtime would allow for rapid deployment across a fleet without
any service disruptions. At the same time, fixing bugs and reiteration
incorporating feedback based on collected telemetry becomes much faster by
unloading and reloading the program at runtime. This significantly simplifies
deployment, accelerates the feedback loop of testing and qualifying kernel
changes, and speeds up development.

\PN{Integration with the Kernel}
While programs attach to the kernel to introduce alternative behavior, they
should be able to interact safely with existing kernel state, and manipulate it
if needed. At the same time, programs should have the option to fallback to the
existing kernel processing in case they have nothing of interest to do. This
presents a flexible model to the user, where the existing kernel implementation
can be made use of, if needed, without having to reimplement similar code within
the program.

\section{Overview}
\label{s:overview}

\begin{figure*}[ht]
	\centering
	\includegraphics*[width=\textwidth]{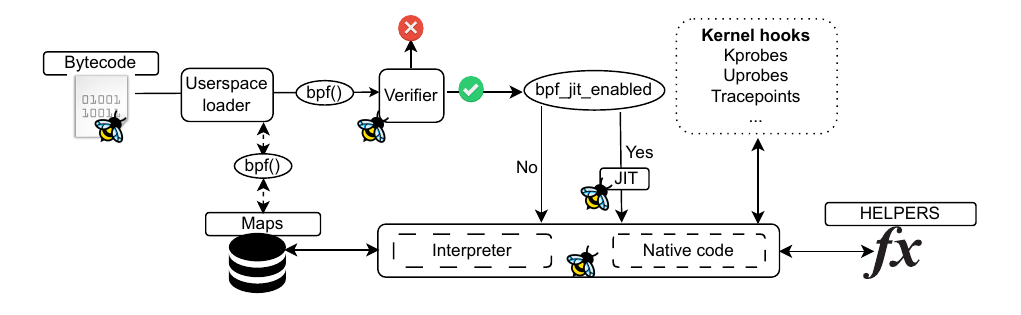}
	\caption{An overview of the \sys key components and their correlation based on \cite{eBPFio_png}.}
	\label{fig:components}
\end{figure*}

\sys is defined as an abstract virtual machine that supports the \sys instruction
set~\cite{insnset}. The virtual machine has a set of 11 registers and a fixed size
stack. An \sys program operates within a restricted virtual machine environment provided by the Linux kernel. The \sys instruction set is a small but versatile collection of 64-bit instructions. These instructions provide a wide range of functionalities, enabling eBPF programs to efficiently perform various tasks within the kernel space~\cite{rice2023learning}. The instruction set supports arithmetic operations such as addition, subtraction, 
multiplication, and bit operations (e.g, \cc{AND,OR,XOR}). Additionally, it supports load and store instruction and jump instructions encompassing both conditional and unconditional jumps to alter the program flow, as well as function calls and exits. The instruction set also features atomic operations designed for safe memory access and modification. These atomic operations ensure that concurrent access does not lead to inconsistencies or corruption.

\subsection{The \sys Runtime}
An \sys runtime is the set of necessary components required to map the abstract
virtual machine and related entities onto another OS or hardware platform. For
the Linux kernel, the \sys subsystem within the kernel implements the
\sys runtime and defines the system call interface to interact with it. 
Below, we present the key components that makes up the \sys ecosystem as shown in \autoref{fig:components}.

\PN{\sys Bytecode}
\sys bytecode is defined as a finite sequence of \sys
instructions. The \sys virtual machine executes \texttt{\sys programs}, which are
encoded using \sys bytecode. Each program is composed of one or more
\texttt{subprograms} (or subprogs in short). These are simply self-contained units of
bytecode analogous to functions. Execution of a program begins at the
\texttt{main} subprog.

\PN{\sys Userspace Loader}
\sys has a vibrant ecosystem of user space loaders such as BCC~\cite{bcc}, Bptrace~\cite{bpftrace} and libbpf~\cite{libbpf} that loads \sys bytecode into the kernel using the \cc{BPF\_PROG\_LOAD} system call and attaches the program to relevant hooks while managing any corresponding maps~\cite{eBPFruntime}. A file descriptor is returned, which represents the program that has been loaded, which can then be used to attach the program to specific kernel hooks. For the purposes of this paper, we keep
the discussion and expositions centered around the LLVM toolchain, its C
frontend, and libbpf, which together remain the most popular and
featureful reference implementations\footnote{These tools are the standard tools that have gained popularity in modern development environments, especially in systems programming and applications that emphasize performance.}

\PN{\sys Verifier}
\label{ssss:ver}
The verifier, a crucial component in eBPF systems, inspects the bytecode before it is accepted into the kernel, ensuring safety properties associated with \sys and that the loading of programs does not negatively impact the kernel's integrity and safety under any circumstances.

\PN{\sys Just-In-Time Compiler and Interpreter}
Upon completion of the verification process, the program is compiled into native machine instructions using the Just-In-Time
(JIT) compiler. The kernel then assigns a file descriptor to the
loaded program, simplifying its attachment to various execution hooks within the kernel. In cases where JIT is disabled
or unsupported, the \sys interpreter takes on the responsibility of program execution, dynamically decoding and executing the bytecode in real time.
\begin{figure*}[ht!]
	\centering
	\includegraphics*[width=0.8\textwidth]{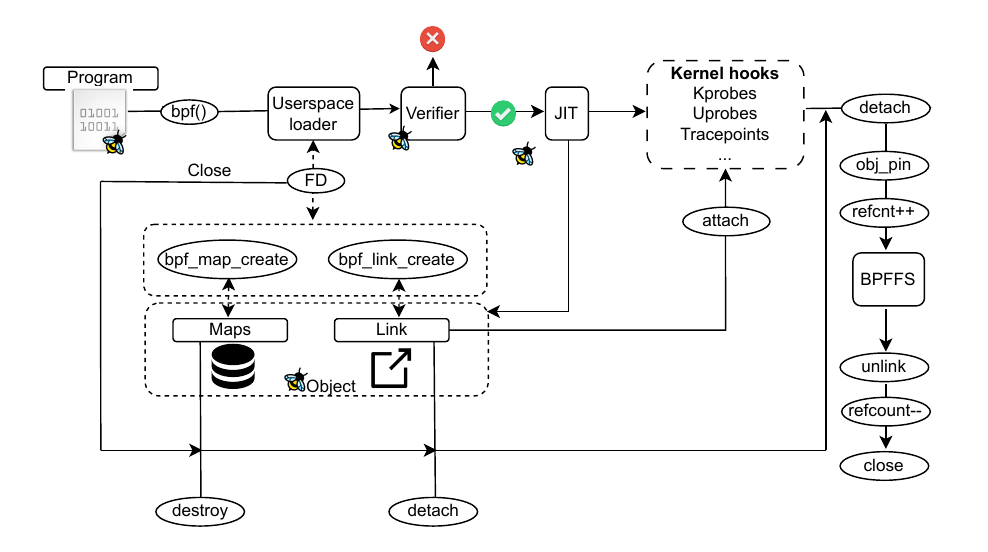}
	\caption{This diagram depicts the lifecycle of \sys objects within the kernel, with the in-kernel representation, interaction with file descriptors, and the role of pinning in the \emph{bpffs} file system \cite{CiliumPinning}.}
	\label{fig:lifecycle}
\end{figure*}

\PN{\sys Hooks}
The \sys program's flow is determined by events, which are executed when the kernel or an application encounters specific hook points. These predefined hook points are placed in various locations within the kernel and cover a wide range of events, including system calls, function entry and exit, network sockets, tracepoints, and more. Based on its attach and/or program type, the program is attached to the hook where it is supposed to run. If a predefined hook does not meet the requirement, developers can create custom hook points called kernel probes (kprobes) or user probes (uprobes). These probes enable the attachment of eBPF programs to almost any position inside the kernel or user applications.

\PN{\sys Program Types}
The \cc{BPF\_PROG\_TYPE} is the categorization of an eBPF program that determines its function, input parameters, acceptable actions, and attach points in the kernel. Each program type has characteristics that define its behavior and interaction with the system. For example, the socket filter program type \cc{BPF\_PROG\_TYPE\_SOCKET\_FILTER} is designed to examine and manage network packets at the socket level. Developers can design customized logic within these programs to analyze and modify incoming and outgoing packets as needed. Conversely, tracing program types \cc{BPF\_PROG\_TYPE\_TRACING} are capable of monitoring kernel events and providing valuable information about system operation. These types of programs and their attach types are defined in the kernel codebase and serve as blueprints for creating eBPF programs to meet different requirements.

\PN{\sys Helpers}
\sys helpers are specialized functions accessible to eBPF programs, enabling interaction with the system and their execution context. These helpers facilitate tasks such as printing debugging messages, retrieving system boot time, manipulating network packets, and interacting with eBPF maps. Each eBPF program type can access a specific subset of these helpers, tailored to its context and requirements. For details about bpf helpers, see the documentation provided by the kernel~\cite{helpers}.

\PN{\sys Maps}
An \sys map is an abstract data structure of a certain type, such as an array or hash map that facilitates data exchange between the user space and the kernel. The programs running within the \sys virtual machine may obtain
access to one or more maps through platform-specific load instructions.

\subsection{\sys Objects and their Lifecycle}
Each \sys object has an in-kernel representation for the management of eBPF program within the kernel, and is exposed through a file descriptor to user space. The lifecycle
of the \sys object is tied to the lifecycle of the file descriptor~\cite{off_ebpf}. Once the
final file descriptor corresponding to the \sys object is released, its state
inside the kernel is also released. To enable persistence beyond a process's
lifetime, the kernel allows \textit{pinning} these file descriptors on a special
pseudo-file system called \emph{bpffs}. Each pinning operation takes a
a reference on the \sys object, thus extending its lifetime.

\PN{\sys Programs}
These objects represent the actual program being loaded into the kernel. A file
descriptor representing the program is returned from the \cc{BPF\_PROG\_LOAD}
command of the \cc{bpf} system call, after the \sys verifier verifies and JITs
the program and creates its in-kernel representation. Once loaded, the program
is ready to be attached to a designated kernel hook.

\PN{\sys Maps}
When \sys programs are created, maps are defined using \cc{BPF\_MAP\_CREATE} command of the \cc{bpf}
system call, which returns a file descriptor. This descriptor is used in pseudo load instructions within the \sys program to reference the map. The verifier resolves the file descriptor to the actual map object
in the kernel during program verification, and treats the destination register
of the pseudo load instruction as a \sys map pointer in the program.

\PN{\sys Links}

\sys links play a crucial role in ensuring that \sys probes outlive the lifecycle of the application triggering them. \sys links are created using the \cc{BPF\_LINK\_CREATE} command
of the \cc{bpf} system call. The link enables developers to indirectly attach \sys programs to kernel hooks providing a more flexible and sustainable method than regular direct attachment methods~\cite{bpflink_syscall}. Instead of attaching directly to a hook, creating an \sys link ties the lifetime of a program's
attachment to a file descriptor, thus simplifying the management of program references and maintaining the probe even if the application loading it terminates unexpectedly~\cite{Andriibpflink}. The file descriptor associated with an \sys link controls its lifecycle.  When the last file descriptor is closed, the link detaches its program from the kernel hook, allowing for resource cleanup. Only the link owner can detach or update it, ensuring system integrity and preventing unauthorized modifications~\cite{Andriibpflinkcgroup}.

\PN{BTF} BTF objects represent the BTF type information for a \sys program or
map which has been submitted into the kernel from user space, to allow the
verifier to tie this type information to the program or map during its
verification procedure. In the case of the kernel and its modules, these objects are
automatically created during boot and when any of the kernel modules are loaded. We elaborate on how BTF has been instrumental in enabling several other use cases in \autoref{ss:btf}.

\subsection{\sys Instruction Set}
\label{ss:insnset}
The \sys instruction set is defined in terms of the \sys virtual machine. It
supports two types of instruction encoding (64-bit and 128-bit),
general-purpose instructions (such as arithmetic, jump, call, load, and store), and
addressing the set of 11 64-bit registers (\cc{r0}-\cc{r10}) with a well-defined
calling convention, where \cc{r10} is read-only and points to the top of the
stack. We defer to the eBPF Instruction Set Specification~\cite{insnset} for the
complete formal description of the instruction set and the calling convention.

A key principle used throughout the instruction set's design is maintaining
near-equivalence with actual hardware instruction set architectures (ISA), which simplifies the implementation of interpreters and JITs.  Moreover, this near-equivalence allows for optimizing compiler backends to emit \sys assembly whose performance is close to natively compiled programs. This is because JITs can translate \sys instructions to
native machine instructions mostly with a one-on-one mapping, without
introducing any extra instrumentation to handle the translation.

One of the strengths of decoupling the instruction set from the \sys runtime in
the Linux kernel is the ability to use it outside of the operating system. The
\sys ISA can either be supported directly by hardware
\cite{githubGitHubRprinz08hBPF}, or translated to the target architecture of the
hardware.  This also has immense potential for computational hardware, as the
\sys runtime can control and decide whether offloading should be performed,
while programs for a certain hook may be written in an oblivious fashion.
Kicinski \etal ~\cite{kicinski2016ebpf,kicinski2017ebpf} have demonstrated the
offloading of \sys programs for the XDP hook~\cite{01Toke2018} to programmable
NICs, while Lukken \etal ~\cite{lukken2021zcsd} explored its use for
computational storage devices.

\subsection{BPF Type Format}
\label{ss:btf}
The BPF Type Format, or BTF for short, is a debug information format designed specifically for use with \sys. It is produced by the compiler when the kernel or an \sys program is
compiled.
In addition to information about C types, it contains function prototype
information, custom annotations for types and declarations, which carry
context-sensitive meaning for the \sys verifier, and source information for
better introspection and debugging. We defer to the BTF documentation~\cite{btf}
for the formal description of the metadata format.

The benefits of a new debug information format are twofold. Firstly, the
existing debug information format used by the kernel, Debugging with arbitrary record formats (DWARF)~\cite{dwarf}, has a
large overhead in terms of memory consumption if embedded in the kernel image.
This means that always shipping kernels with debug information which could be
used by the \sys verifier to enrich its static analysis was infeasible.
Secondly, for better introspection and analysis of \sys programs and maps, they
would need to supply their own debug information which could be inspected by the
verifier. This meant introducing complex code to parse DWARF debug information
into the kernel, which was undesirable for maintenance and security reasons.

BTF addresses all of this concern. Due to its compact representation,
there is an order of magnitude difference~\cite{btfdedup} in the memory
consumption between DWARF and BTF debug information for the same kernel image
produced by the compiler. This is primarily due to the aggressive deduplication
algorithm devised by Nakryiko \etal~\cite{btfdedup} to reduce BTF's memory
footprint.

In turn, this allows the BTF to always be shipped with the kernel and \sys
programs, which the verifier heavily relies on to perform its static analysis.
Due to its simpler representation, BTF is also faster to process, which is
critical as an in-kernel representation is created at runtime for the kernel,
any loaded kernel modules, and all \sys programs and maps supplying their BTF
information.

We now illustrate BTF's advantages in the context of its primary use cases.

\PN{Verification} The main consumer of BTF is the \sys verifier. It uses the
kernel's BTF information to enforce \textit{type safety} (\autoref{sss:safety})
in \sys programs for kernel pointers they gain access to. The verifier will
introspect the type information using BTF to ascertain the size of the object and introspect members of a structure type.

\PN{Annotations} BTF can carry custom annotations for types and declarations of
functions and variables. These are used to attach context-sensitive meaning~\cite{rcu}, to types used in the kernel or the program, to aid
verification~\cite{kptrs}.

\PN{Debugging} The use of BTF in \sys programs and maps allows for better
introspection and debuggability. When analyzing a program, the verifier can
print the source and line information to its log, which comes in handy whenever
a program is rejected for user output. It also allows annotating the \sys
bytecode and JIT compiled code with source information. For \sys maps, dumping
of their data can be made structural by recognizing the type of data from their BTF.

\PN{Compile Once Run Everywhere (CO-RE)} 
\label{par:co-re} 
CO-RE is the collective name of a set of relocations for \sys programs. These allow compiled programs to be more portable by encoding symbolic references for memory accesses to members of
structure types, named enumerator constant values and named kernel
configuration options. All of these relocations are resolved either by the
verifier or \cc{libbpf} when loading the program. This dynamic resolution ensures that eBPF programs can adapt to different kernel versions and architectures without the need for recompilation.

\section{Workflow}
\label{s:workflow}
\begin{figure*}[t!]
	\centering
	\includegraphics*[width=0.9\textwidth]{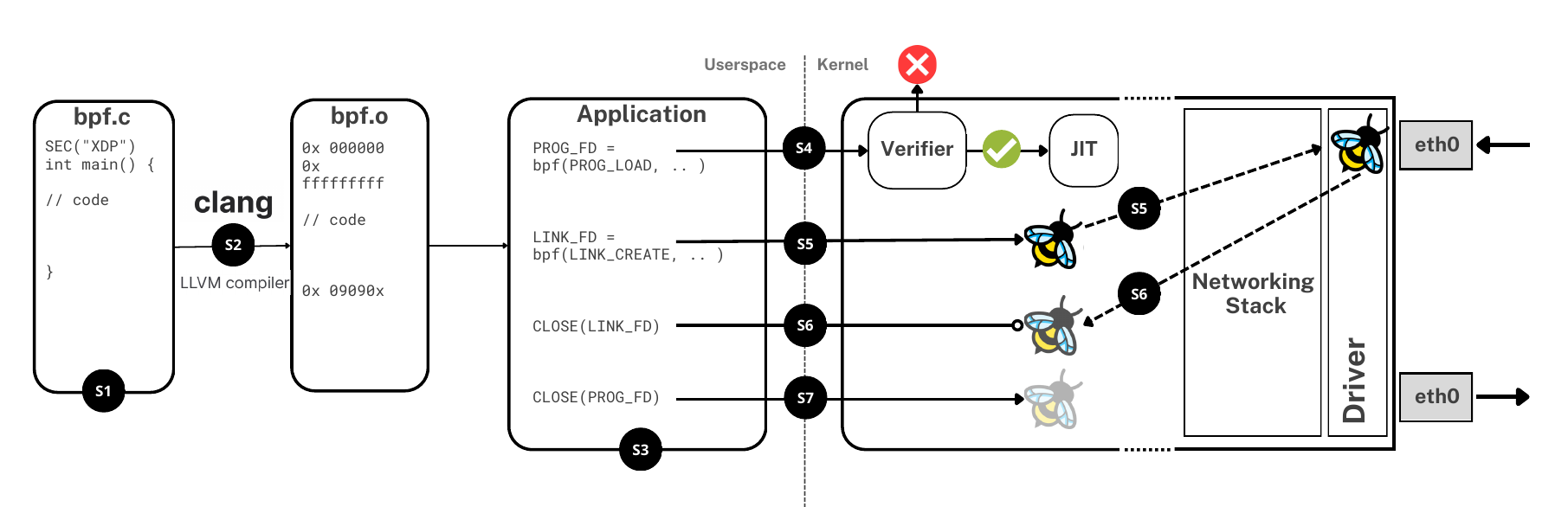}
	\caption{Workflow diagram of an eBPF program based on \cite{CiliumPinning,eBPF_workflow}.}
	\label{fig:workflow}
\end{figure*}

In this section, we illustrate the high-level programming and execution model of
\sys. \autoref{fig:workflow} illustrates the entire sequence involved in the
process of writing and executing an \sys program for our chosen example from
start to finish.

A user typically begins at step \BC{S1} by authoring an \sys program in a
high-level programming language. For our example, we consider the C program in
\autoref{lst:xdp} written for the XDP hook \cite{01Toke2018}, which invokes the
program for network packets at the network device driver layer before they are
processed by the networking stack. The \cc{ctx} argument of type \cc{struct
xdp\_md} represents the raw network packet the program gets access to. The
\cc{data} and \cc{data\_end} pointer variables point to start and one past the
end of the network packet's data area. Conditional branches comparing \cc{data}
and \cc{data\_end} are used to ensure that enough room is available, to ensure
that memory accesses of the packet's data will be safe.

\begin{figure}[H]
\centering
\begin{lstlisting}[language=c, caption={An example BPF program for the XDP hook, which drops all incoming IPv4 UDP traffic.}, label={lst:xdp}]
SEC("xdp")
int bpf_program(struct xdp_md *ctx)
{
  void *data_end = (void *)(long)ctx->data_end;
  void *data = (void *)(long)ctx->data;
  struct ethhdr *eth = data;

  if (eth + 1 < data_end) {
    if (eth->h_proto == bpf_htons(ETH_P_IP)) {
      struct iphdr *iph = (void *)(eth + 1);
      if (iph + 1 < data_end && iph->protocol == IPPROTO_UDP)
        return XDP_DROP;
    }
  }
  return XDP_PASS;
}
\end{lstlisting}
\end{figure}

The next step \BC{S2} involves the compilation of this C program using the LLVM
toolchain's clang compiler into an object file. The target for compilation is
chosen as \cc{bpf}, which instructs the compiler to use the \sys backend to emit
binary code for the produced object file.

Step \BC{S3} is concerned with the processing of the produced object file and
submitting the program encoded within it to the kernel through the \cc{bpf(2)} \cite{bpf_syscall} system call for loading it. For our example, we use the \cc{bpftool}
user space tool which in turn uses the \cc{libbpf} library to perform the loading
of the program. Once the object file has been processed and the program has been
extracted from it, step \BC{S4} submits it to the kernel using the \cc{bpf(2)}
system call's \cc{BPF\_PROG\_LOAD} command, which invokes the \sys verifier.

The \sys verifier then performs verification of the program to decide whether it
will be safe for execution within the kernel. If the verifier fails to determine
the program's safety, it rejects the program and returns an error to user space.
Otherwise, a successfully verified program is JIT compiled, and a file
descriptor corresponding to the \sys program is returned to user space.

Once user space has the file descriptor for the program, it can now attach to a
network device's XDP hook. For our example, the network device is \cc{eth0}, and
step \BC{S5} involves invoking the \cc{BPF\_LINK\_CREATE} command of the
\cc{bpf(2)} system call to perform the attachment of the program to the network
device. If all parameters for the command were valid, the kernel returns a file
descriptor corresponding to the \sys link is returned to user space.

At this point, the \sys program is attached to the \cc{eth0} network interface and
rejects all incoming IPv4 UDP traffic for it. It is invoked for every raw
network packet received by the kernel's network device driver, and performs its
processing on it. The rest of the traffic passes through to the kernel's
networking stack.

In step \BC{S6}, once the user space application closes the \sys link
file descriptor, the program is detached from the network interface. Once the
\sys program file descriptor is closed in step \BC{S7}, the kernel will free the resources it
occupies such as memory for its code.
\section{Safety of \sys Program}
\label{s:safety}
Program safety is a critical aspect of \sys programs, ensuring they execute correctly and securely without compromising the stability and security of the Linux kernel. In the context of \sys program, program safety refers to a set of properties that must be satisfied to protect the runtime integrity of the kernel and uphold the invariants of the kernel context in which the eBPF program will execute. Programs that violate any of these safety properties should be rejected when being loaded into the kernel.

\subsection{Safety Properties}
\label{sss:safety}
We now classify and discuss each safety property as detailed in the BPF verifier code and related documentation~\cite{bpf_verifier, verifier_docs}.

\begin{itemize}
    \item \textbf{Memory Safety:} For all memory regions accessible to the program, the verifier endeavors to guarantee that there can be no out-of-bounds accesses, invalid or
    arbitrary memory accesses, or use-after-free errors. The verifier performs precise bounds tracking for all memory regions accessible to the program and checks that all accesses lie within bounds. Any allocated memory region cannot be accessed once it has been freed. Arbitrary values cannot be used as pointers for memory accesses. 

    \item \textbf{Type Safety:} The verifier has precise knowledge of the type of each register
    that points to a memory region accessible to the program. It keeps track of the
    type of objects on the stack. This process helps avoid errors caused by type confusion and ensures that different types of programs can utilize various utilities without corrupting kernel memory. The verifier also leverages BTF (\autoref{ss:btf}) to capture essential information about kernel and eBPF program types and code. BTF helps identify eBPF kernel data structures and ensures that aggregate types, such as structs, are not accessed beyond their allowable limits.

    \item \textbf{Resource Safety:} The verifier checks that the program leaves no lingering
    resources when it exits. This implies that the program must release all
    allocated memory, acquired locks, and reference counts to kernel objects using
    an appropriate helper function.

    \item \textbf{Information Leak Safety:} 
    The eBPF verifier outlaws any kernel information leaks by analyzing pointers that could potentially reference kernel memory, aiming to prevent any leaks into user-accessible memory regions. It performs thorough escape analysis~\cite{escape_analysis} for all pointers that may point to kernel memory, ensuring they do not escape into memory regions accessible by user space. Additionally, it rejects any attempts to read uninitialized regions of the stack.

    \item \textbf{Data Race Freedom:} The verifier aims to ensure that the program's accesses to kernel state are free from data races. It enforces that any manipulation of
    kernel state occurs through helpers implementing appropriate synchronization.
    However, the verifier does not diagnose data races for accesses to memory owned
    by the program itself (\eg~values of \sys maps), because it has no bearing on
    the kernel's runtime integrity.

    \item \textbf{Termination:} The verifier places a limit on the maximum number of instructions it will explore across all paths of a program, known as the \texttt{instruction complexity limit}, aiming to ensure that all programs terminate. If the program does not demonstrate termination for all paths within this limit, the verifier rejects it. This could be due to infinite loops, unbounded loops with unproven termination, or simply because of programs that are too large.

    \item \textbf{Deadlock Freedom:}
    The verifier aims at ensuring that the program is free of deadlocks. By definition, a
    deadlock requires two programs executing in parallel to hold two locks in the
    opposite order. To achieve deadlock freedom, the verifier simply disallows
    holding more than one lock at a time at any point in the program.

    \item \textbf{Upholding Execution Context Invariants:}
    The verifier checks to ensure that the program upholds all invariants of the execution
    context within the kernel. In other words, a program's execution may not violate
    the invariants and assumptions of existing kernel   code. This information is
    encoded in the verifier every time support for a new kernel hook is introduced.
\end{itemize}

By delineating these safety properties, we have defined the state of the art for what the verifier will enforce for \sys programs. However, it is crucial to recognize that these standards are not static; rather an ongoing work to protect the integrity and security of \sys programs within the kernel, and will be periodically updated as the ecosystem changes and new approaches are explored to reflect the dynamic nature of this field.

\subsection{The \sys Verifier}
\label{ss:verifier}
To enforce the safety properties (\autoref{sss:safety}), the \sys verifier plays a major role. The \sys verifier is a static analyzer for \sys programs. It is invoked when a
program is loaded into the kernel, and is tasked with ensuring that the program
is \emph{safe} to execute within the kernel's context. Once this has been
determined, the verifier submits the bytecode for JIT compilation, where
it is converted into native machine instructions. The program can then
be attached to one of the many available kernel hooks to begin its execution.

The verifier performs static analysis of the program at the level of its \sys
bytecode. It has access to the program's BTF (\autoref{ss:btf}) debug
information which was produced during its compilation from a high-level language
to \sys bytecode.
This design choice of operating over \sys bytecode and BTF allows the \sys
verifier to be usable for programs written in multiple higher-level languages.

\begin{figure}[t!]
	\centering
	\includegraphics*[width=\columnwidth]{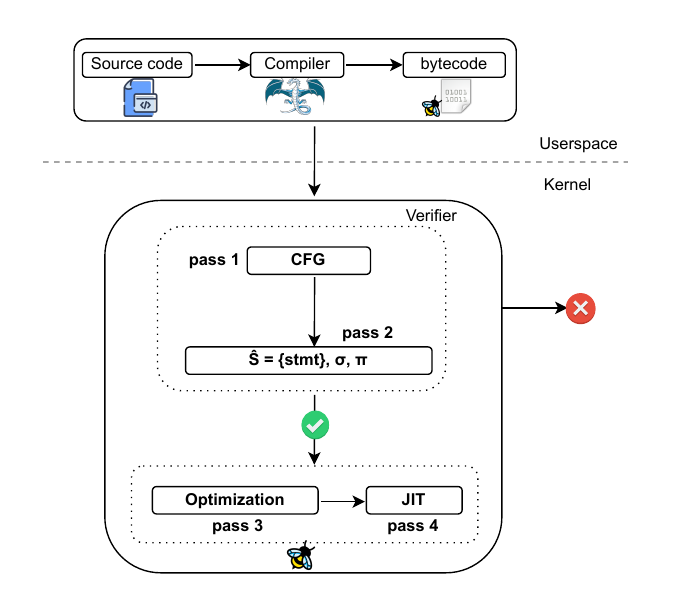}
	\caption{The four major passes of the \sys verification process}
	\label{fig:verifier}
\end{figure}

There are four major passes involved during the verification of a program, as shown in \autoref{fig:verifier}. The
first pass validates the control-flow graph of the program
(\autoref{ss:cfg}). The second pass performs exhaustive symbolic execution of
the program to ensure that it is \emph{safe} (\autoref{ss:symbex}). Thereafter, the third pass performs
optimizations and transformations (\autoref{ss:opt}) for the program before it
is submitted to the Just-In-Time compiler (\autoref{ss:jit}) in the final pass. The following sections describe the verifier passes, explaining how the verifier enforces the safety properties.

\section{Validation of the Control Flow Graph}
\label{ss:cfg}
The verifier begins by analyzing the Control-Flow Graph (CFG) of the program.
Instructions in the CFG are represented as nodes, and the different
forms of control flow represent edges. All, but the first instruction, will
have at least one incoming control-flow edge. Except \cc{EXIT}, all instructions also have
one outgoing \texttt{fallthrough} control-flow edge, which points to the next
instruction. For unconditional \cc{JUMP} instructions, the fallthrough edge
points to the instruction they jump to.
In case of \cc{CALL} and conditional \cc{JUMP} instructions, there is an extra
outgoing \emph{branch} control-flow edge to the branch target. For \cc{CALL}
instructions, the branch control-flow edge points to the callee's first
instruction. For conditional \cc{JUMP} instructions, it points to their branch
target instruction (when the condition is true).

The verifier walks the CFG (\autoref{s:appendix-fig:cfg}) of the program using a depth-first search algorithm.
This allows it to label all reachable instructions and the control-flow edges
followed to visit them. Simply completing the traversal allows it to detect any
unreachable instructions in the program by checking if they were unvisited. Additionally, while walking the CFG, it tracks \texttt{state pruning} points.

The verifier checks for the following properties, and rejects programs that do not adhere to it:
\begin{itemize}
	\item{Ensures no infinite loops or loops with complex termination conditions that cannot be statically determined.}
	\item{Ensure that there exist no unreachable instructions in the program.}
	\item{Ensure that all subprogs end with \cc{EXIT} or
	      \cc{JUMP} instruction, \ie~no automatic fallthrough to the next
	      subprog.}
\end{itemize}

\PN{State Pruning.}
\label{sss:pruning}
The verifier performs precise symbolic execution of the \sys program (described in \autoref{ss:symbex}). However, this approach can be very expensive when multiple paths need to be explored due to the presence of multiple branch conditions. The distinct state constraints, symbolic and concrete values for registers and the stack have to be maintained in separate verifier states.  Thus, the path explosion problem also translates into a state explosion problem. Since the verification algorithm's complexity limit concerns instructions explored in \textit{all} possible paths, larger and more complicated programs would eagerly hit this limit.

To alleviate such scalability issues, the \sys verifier implements the
\texttt{state pruning} approach, borrowing ideas from the RWSet analysis
technique by Boonstoppel \etal~\cite{boonstoppel}. This involves pruning of
redundant path walks while exploring the program by detecting variable
\emph{liveness}. The idea is to establish equivalence in terms of program
side-effects between the current program state and an already explored program
state which has passed through the same point in the program. During this state
equivalence check, the verifier skip variables which are not used in
subsequent basic blocks. These pruning points are simply instructions in the program for which the states will be stored and compared for equivalence every time the verifier encounters them while exploring a path.

The program is annotated with \texttt{pruning points} where the verifier's current state will be saved. Later, when the verifier arrives at the same pruning point while exploring another path, it traverses through the checkpointed states which have been fully explored and
compares the current state to them.

This is meant to establish \texttt{state equivalence}, a property which takes
into consideration whether given the old checkpointed state, which has already
been verified to be safe, it can be considered equivalent to the current state
from the point of view of program safety. If this property is true, then the
verifier need not continue exploration of the current path, as the program from
this point has already been verified to be safe for an equivalent verifier
state. Otherwise, the verifier keeps exploring the current path until the next
pruning point, where it repeats this step, or it encounters the \cc{BPF\_EXIT}
instruction.

\section{Symbolic Execution}
\label{ss:symbex}
The verifier's second pass symbolically executes the \sys bytecode. To ensure
safety, as defined previously in \autoref{sss:safety}, the verifier must
exhaustively explore all feasible paths through the program and precisely track
state of the stack (at byte-level granularity) and the program registers through every point in the the control
flow. Note that pass one (\autoref{ss:cfg}) already ensured that all points in
the CFG are reachable.

State tracking in \textit{symbolic execution} is defined in the form of a three-tuple $(\hat{S} = \{stmt\}, \sigma, \pi)$ 
for every path explored.
$\{stmt\}$ denotes the list of instructions visited while executing the current path symbolically.
$\sigma$ is a symbolic store that maps \emph{program variables}, here \sys
registers and stack, to symbolic or concrete values and tracks associated type
information. Finally, $\pi$ represents the path state, which includes additional state information maintained during path exploration. In the context of \sys, these conditions maintain dataflow information collected during path exploration.
For example, $\pi$ maintains \textit{precision} markings to know which variables need to be precisely tracked (\autoref{sss:pruning}), lock-held sections of the program. 
In addition to tracking this three-tuple over all paths, the verifier also caches the state $(\sigma, \pi)$,
by checkpointing them at \textit{prune points} that we computed in pass one (\autoref{ss:cfg}).
We call this as the \textit{verifier state} in
the context of the \sys verifier's symbolic execution pass.

The verifier begins path exploration from the first instruction of a program, by
setting up the \textit{symbolic store}. In the initial state, only register $r1$
has some value. It holds a pointer to the context object which every program
receives when being invoked from the kernel; all other registers are empty and the stack is uninitialized.
The internal structure of the object
pointed to by the context pointer is dependent on the program type. Thereafter,
the verifier begins symbolically executing the program, instruction by
instruction, and continues to update the \textit{verifier state}.

When the verifier encounters a conditional \cc{JUMP} instruction, it first tries
to predict which branch will be taken based on the values of operand registers
in the symbolic store $\sigma$. If it cannot predict the evaluation of the
branch condition, the verifier splits exploration by forking the current
\textit{verifier state} into two. It then continues exploration of one path of
the branch, and enqueues the other path into a worklist for later exploration.
The \textit{verifier state} in each path exploration will have additional
information about the program state based on the truth assignment of the branch
condition. The symbolic value of the operand register will be updated to have
more precision. 

Unfortunately, with branching programs such as loops, the number of
paths to explore increases exponentially, triggering path explosion. To keep the
verification task tractable, the verifier employs an approach called
\textit{state pruning} which reduces the number of paths it needs to explore. As
we discussed earlier in \autoref{sss:pruning}, this ties back to the \textit{prune}
points marked by  the verifier during its CFG validation pass. We now
discuss in detail the pieces needed for symbolic execution.

\subsection{Symbolic State~}
The symbolic store $\sigma$ in the \textit{verifier state} tracks the following
information about registers and stack state:
\begin{itemize}
	\item It tracks whether the value in a register, or data in the stack (at byte-granularity), is a \textit{scalar} or a \textit{pointer}.
	\item For \textit{scalars}, the verifier tracks precise integer bounds. This means that the verifier does not over-approximate the range based on visiting a $stmt$.
 	\item For \textit{pointers}, the verifier tracks the type of object it points to, the \cc{offset} of the pointer in an accessible memory region, and the \cc{len} in which de-referencing the pointer is valid.
\end{itemize}

The path state $\pi$ in the verifier tracks the following information at every \textit{prune} point:
\begin{itemize}
    \item \textbf{Liveness Tracking:} The verifier performs, in-place, live variable analysis to backpropagate which registers are live.
 
    \item \textbf{Scalar Precision:} In addition to liveness, the verifier also tracks scalar values that are used later in contexts where precision tracking is necessary, such as \cc{JUMP} targets and helper calls. Registers that hold scalars that are live but need not be precise also assist in pruning redundant path exploration.
 
    \item \textbf{Resource Tracking:} $\pi$ also keeps track of resources like allocated memory regions, ref-counts and spinlock held in a certain section of the program to ensure their timely release before program termination. It enables the management of various mechanisms including lock tracking and synchronization, kernel object reference management, and memory allocation and deallocation.
 
    \item \textbf{Pointer Alignment:} The verifier inspects the alignment of register types and pointers to prevent any access to memory with incorrect alignments due to variable offsets. It scrutinizes the amount of data being read from or written to memory in a single operation. Additionally, it enforces alignment checks for stack pointers, as it depends on tracking stack spills. Any misaligned stack access can lead to the corruption of spill registers, thereby posing potential threats of exploitation.
\end{itemize}

\subsection{Instruction Simulation}
The \sys instruction set defines all instruction types, their expected operands, and expected behavior.
All instructions are encoded into the verifier as transfer functions that take an initial state $\hat{S}_{0}$ and returns a new state $\hat{S}_{1}$. If any simulated transition causes an error, verification fails.

We now discuss what the modeling of each instruction-class looks like in the verifier.

\PN{Arithmetic Operations}
In cases where the opcode is valid, the function goes one step further and
checks the validity of the source and destination operands by ensuring that the
register used as the source operand is readable and that the register used as
the destination operand can be written to. If the source and destination
registers are valid, arithmetic operations with pointers and scalars are
processed and new signed and unsigned bounds can be calculated for all
\emph{32/64-bit} ALU operations, excluding \cc{BPF\_END}, \cc{BPF\_NEG} and
\cc{BPF\_MOV} operations as detailed in \autoref{s:appendix-ebpf_ops}.

\PN{Load and Store Instructions}
Store instructions have two modes of operation. The \cc{BPF\_STX} class takes a
register operand whose value will be stored, while \cc{BPF\_ST} works with an
immediate as the source operand. The destination operand is always a register
with a pointer type of a memory region. While the handling of each pointer type
is distinct, overall all of them share a few high level properties. First of
all, the actual offset where the store is done is formed by accumulating the
offset specified as part of the store instruction and the symbolic or concrete
offset value associated with the destination register.

\PN{Value Tracking} 
Value tracking ensures the safety and accuracy of memory operations in eBPF programs by analyzing register and stack slot values and enforcing constraints. Each register's state is carefully observed and assigned a specific type based on its content. For instance, registers may have type \cc{NOT\_INIT} if they haven't been written to yet, or \cc{SCALAR\_VALUE} if they hold scalar values not used as pointers. Pointers, when present, are categorized into various types like \cc{PTR\_TO\_CTX} or \cc{PTR\_TO\_MAP\_VALUE}. Additionally, pointers can undergo arithmetic operations, resulting in fixed or variable offsets. The verifier tracks these offsets, adjusting their minimum and maximum values accordingly. This insight of register types and pointer offsets ensures that memory accesses within eBPF programs adhere to designated areas and comply with safety constraints.

\PN{Numerical Abstract Domain}
\label{sss:tnum}
Tnum is an abstract range that approximates the set of possible values that can
be stored by a variable. It is used by the \sys verifier to check the validity of
memory usage during read or write operations. Tnum is a 64-bit value with a mask
that indicates a range of possible values. Each bit in the binary representation
of a tnum can be set to 1, 0, or unknown~\cite{tnumvish}, representing a precise
value, a known range, or an unknown range, respectively. The verifier also
tracks register constants using an interval abstraction to determine the minimum
and maximum possible values of each register at any given time. By analyzing
this data, the verifier can detect potential out-of-bounds memory accesses and
prevent them from occurring during program execution.
For the complete formal description of the verifier's numerical abstract domain,
including the proofs for its soundness and optimality, we defer to Vishwanathan
\etal~\cite{tnumvish}.

\PN{Conditional and Unconditional Jumps}
In the eBPF architecture, jumps are classified into unconditional and conditional. Unconditional jumps move the program counter to a new instruction by adding a fixed offset to the current instruction's position. This offset can be positive or negative, enabling jumps forward or backward, as long as they stay within the program's boundaries and avoid loops.

Conditional jumps, however, depend on a runtime condition involving the evaluation of certain variables or registers. If the condition evaluates to true, the program counter is adjusted by a specific offset to reach a new instruction. If the condition is false, the execution continues to the next sequential instruction. Thus, unconditional jumps follow a predetermined path based on a fixed offset, while conditional jumps adjust the path according to whether the evaluated condition is met or not\cite{bpfandxdpref}.

\PN{Function Calls}
Different eBPF program types have access to a different set of functions, reflecting their specific use cases. Function calls in eBPF programs extend their functionality. To ensure safety, the verifier enforces strict checks to guarantee that functions are called with valid arguments.This process involves checking that the registers used for function arguments such as \cc({r1 - r5}) match the expected types, only calls to known functions are allowed and function calls and dynamic linking are prohibited~\cite{function_calls}.

Registers used temporarily within functions, known as caller-saved or volatile registers, need not retain their values after function calls. Therefore, it is the caller’s responsibility to save these values if they are needed later. In contrast, callee-saved registers \cc{r6 - r9} must be preserved across function calls to maintain their values. The verifier’s symbolic execution pass ensures these rules are followed, enhancing function call safety.

The verifier examines whether the program's state has sufficient precision to prove the need to avoid further unrolling. This refers to accurately capturing the critical parts of the state necessary to confirm the program's safety. When precision is achieved, it means the verifier has a stable and detailed understanding of the program state, particularly the aspects that affect safety-sensitive operations. In this case, further actions like unrolling loops are unnecessary. Unrolling is generally used to investigate different possible behaviors of the program, but if the verifier has already obtained sufficient information to ensure safety, it can skip this process.

\subsection{Loops}
\label{sss:loops}
Loops are supported by the verifier in primarily two flavors. The first is loops
within the program whose bounds are known to the verifier. The second is when
the verifier encounters a loop whose bounds are unknown to it.

\PN{Bounded Loops}
The process of bounded loop verification is done by unrolling the loop. The
verifier continues unrolling until it has exhausted exploring all iterations, or
until the instruction complexity limit is hit. Hence, there are no issues
stemming from unrolling until a fixed bound which may be less than the actual
bound, and the verifier does not have to worry about lost precision from not
exploring the remaining iterations.

\PN{Unbounded Loops}
Unbounded loop verification in the verifier relies on special helper functions.
Typically, this is only supported for \cc{bpf\_iter} helpers which initialize an
iterator object, get the next item from the iterator, and destroy it. The
typical setup is to initialize and destroy the iterator object before and after
the loop, and use the return value of the helper returning next item to continue
loop iteration. The verifier knows that the helper eventually returns an item
which terminates iteration, so it does not have to concern itself with proving
the termination of the loop.

At this point, the only relevant property is to ensure that the verifier can
precisely track the state of the program after the body of such loops has
executed an unknown number of times. Due to the limitation of unrolling until a
fixed bound, as premature termination of unrolling may occur, the verifier
employs state pruning logic to establish that the current state already
possesses sufficient precision, thereby proving that further unrolling is
unnecessary. Until the unrolling process converges for such unbounded loops and
performs state pruning at the point where the loop condition is checked (to
determine whether to break or continue iteration), the verifier continues
exploration. If the loop does not converge until the instruction complexity
limit is reached, the program is rejected.

Since the verifier relies on state pruning to establish equivalence of verifier
states representing different iterations of an unbounded loop's body, it needs
to know what registers and stack slots actually need precision to increase the
chances of convergence. To achieve this, the verifier first performs a
depth-first exploration of the remaining program when encountering the loop
condition for the first time. Once it has explored the program past the loop
once,  all precision and liveness marks have been propagated to the checkpointed
states for the relevant pruning points. This then allows the verifier to more
aggressively perform state pruning to establish convergence.

\PN{Path Explosion}
The loop handling in both cases is effective when the loop body is
simple. However, the verifier encounters difficulties when the loop
body contains branches. For each such branch, when the verifier unrolls the loop
body, it has to fork the states to follow both arms of a branch condition. If
one of the paths causes loop termination, it will continue exploring the rest of
the program for each iteration. The number of states grow exponentially when
branch conditions exist within the loop body, and pruning opportunities are
typically missed as each state requires precision for certain registers and
stack slots which are distinct for either of them. This is a major limitation
for the verifier today, where it will simply give up under path explosion once
it hits the instruction complexity limit.

\subsection{Resource Management}
The verifier's resource management logic is concerned with tracking symbolic
resources created for a particular verifier state, and ensuring that they have
been destroyed by the time the exploration of the path being explored
terminates. In simple terms, this means that a \sys helper can be responsible
for acquiring a reference in the verifier state, and another \sys helper will be
responsible for releasing it. This is then mapped to multiple higher-level
acquire-release patterns, such as acquiring and releasing reference counts of
kernel objects, allocating and freeing memory, obtaining and relinquishing
ownership of an object. Mapping verifier level references to values
returned from \sys helpers allows enforcing the invariant that any acquired
resources are released before the program's exit. Each reference's unique
identifier is attached to the register \cc{r0} after a reference acquiring \sys
helper is invoked. The same pointer value is then passed as an argument to a
reference releasing \sys helper, which releases the reference state
corresponding to the identifier. The verifier implements precise tracking of
these references and complains whenever there is an unreleased reference state
after encountering the \cc{BPF\_EXIT} instruction.
Among the resources, the verifier keeps track of held spin locks, and ensures deadlock avoidance. Currently, each program can only hold a single spin lock at once, and must release this same lock before the program ends. By ensuring that every program only holds a
single \sys spin lock at a time, it also eliminates the possibility of
deadlocks. In the verifier state, the verifier associates a unique identifier with the memory region holding the a given \sys spin lock, and remembers this identifier. A pointer to the same memory region must be passed to the unlock function to release the spin lock in the verifier state, thus also ensuring that there is no mismatch between the lock and unlock calls.

\section{Post-Verification Optimizations}
\label{ss:opt}

After completing the symbolic execution pass for the \sys program and
deeming it safe for execution, there are a series of post-verification
optimizations and fixups that are performed on each program as part of the
program optimization and fixup pass. We discuss them in this section.

\subsection{Dead Code Elimination}
The verifier during symbolic execution does not explore untaken branches of
conditional jump instructions when it can determine that a given branch will
never be taken by analyzing the program's data flow. Every instruction that has
been simulated at least once is labeled as `seen'. Thus, once the verifier can
conclude that an instruction is never reachable at runtime, it can safely
eliminate such dead instructions from the program.

Some conditions in the program can be only
resolved as late as the verification stage (\eg, configuration option
values for the kernel on which the program is being loaded).  Thus,
the \sys verifier can eliminate dead code in the program more aggressively than
the compiler due to the richer data flow information available to it.  This
optimization opportunity is readily exploited by the verifier, primarily to
enable better portability for \sys programs across kernel versions, where
different code paths are taken for different versions or configurations, and to
reduce runtime overhead of always untaken branches and conditional jumps.  For
programs running on the kernel's latency critical path (\eg~XDP), dead code elimination is
important for saving precious CPU cycles at runtime due to the nanosecond scale
execution requirements~\cite{01Toke2018} while retaining the ability to have
conditional fallbacks in the program that are resolved at load time.

\subsection{Inlining and Instruction Rewriting}
In some cases, the overhead of direct calls starts to add up for programs. The
underlying operation involves so little work that the overhead of calling a
function to do it would be more than directly executing those instructions. This
also applies to timing-sensitive operations, such as the \cc{bpf\_jiffies64}
helper. However, \sys helpers also serve as a means to enforce API usage and
program context invariants. The verifier cannot permit the program to directly
manipulate kernel data structures. In such cases, it will transparently replace
the helper call instruction and substitute a set of \sys assembly instructions
implementing equivalent functionality. This technique is used for map accesses. During program verification, the verifier knows which \sys map objects the program has access to, statically. Thus, whenever the program calls \sys helpers manipulating those maps (\eg, \cc{bpf\_map\_lookup\_elem},
\cc{bpf\_map\_update\_elem}, and \cc{bpf\_map\_delete\_elem}), the verifier, by
knowing the type of the map during verification, can transparently translate
indirect calls for the map operation made by these \sys helpers into direct
calls for the underlying map implementation. 

\begin{figure*}[htbp!]
	\centering
	\includegraphics[width=\textwidth]{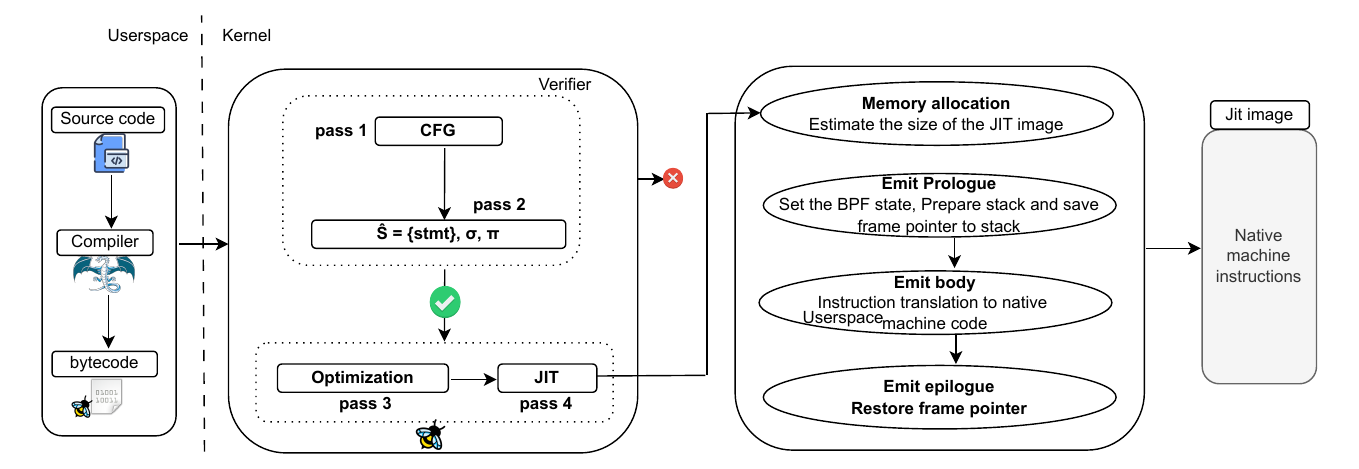}
	\caption{Overview of the process that shows \sys instructions translation into native machine code}
	\label{fig:JIT}
\end{figure*}

\section{Just-In-Time Compilation}
\label{ss:jit}
After after post-verification optimizations, all \sys programs are Just-In-Time
(JIT) compiled to native machine instructions before being enabled for execution. The
JIT compiler performs a direct translation of \sys instructions to the
underlying machine instructions, benefiting from the close equivalence between
the \sys ISA and hardware ISAs. The compilation procedure occurs separately for
each subprog, and can be decomposed into four major steps as illustrated in \autoref{fig:JIT}:
\begin{itemize}
    \item \textbf{Image Allocation:}
Estimate the size of the JIT \textit{image} and allocate memory to write native
machine instructions to after translation.

    \item \textbf{Emit Prologue:}
Emit the code required for the kernel to safely call into the \sys program. This
involves saving the frame pointer and pushing any callee-saved registers to the stack.

    \item \textbf{Emit Body:} Emit the code to translate each \sys instruction to native
machine instructions. Special handling is performed for load instructions which
were seen for registers holding untrusted pointer. The JIT compiler prepares an exception table indexed by the instruction, and any page faults on invalid access at runtime are automatically
resolved to not crash the kernel by looking up this table, and simulating a load of zeroed memory.

    \item \textbf{Emit Epilogue:} Emit the code to undo the prologue, \ie~to restore the frame
pointer and pop the callee-saved registers from the stack.
The final JIT image is made read-only as a security measure to prevent modifications of the
memory later.
\end{itemize}

When extra hardening is requested~\cite{jitguard}, the JIT compiler
performs \emph{constant blinding} as a mitigation. For every instruction using an immediate as
operand, the immediate value is xored with a random constant, and moved into a
JIT-specific internal \sys register (\cc{rAX}). Then, \cc{rAX} is xored
again with the same constant~\cite{Hardening,athanasakis2015devil}, and the instruction in question is translated but
modified to use register \cc{rAX} as its operand. This changes the immediate
value in the produced JIT image after translation, rendering any JIT spraying
attempts useless~\cite{berlakovich2022look}.

\section{Use Cases}
\label{s:usecases}

We illustrate the various use cases served by \sys within and beyond the
Linux kernel.

\subsection{Networking}
We now highlight several key functionalities enabled by \sys programs in networking, empowering developers to achieve high-performance tasks, ensure flexibility, and enhance efficiency through customizable hooks in the Linux kernel.

\begin{itemize}
    \item \textbf{XDP and TC:}
Programs for the XDP~\cite{01Toke2018} and TC hooks~\cite{borkmann2016getting,
borkmann2016advanced, netschedclsact} perform high-performance network packet
processing by bypassing the kernel networking stack, while still having access
to its state. This includes dropping packets at high rates for DDoS
mitigation~\cite{bertin2017xdp,xdp_ddos}, load balancing, implementing network
functions~\cite{ebpf_network_func}, redirecting flows to different CPUs or network devices, or
rewriting and transmitting the packet directly without passing through the
kernel.

     \item \textbf{Socket Lookup:} This hook allows programs to select the socket which
receives a network packet destined for local delivery~\cite{BPFsklookup,
Itscrowd}. This allows steering traffic from any IP address and port pair to a
single socket, without using a socket per pair, which affects socket lookup
scalability.

     \item \textbf{Socket Reuseport:} This hook allows programs to choose one from a set of
reuse sockets bound to the same IP address and port. Programmability through
\sys allows for more informed selection, such as favoring NUMA locality,
connection migrations from one socket to another~\cite{Lameter2013}, or based on data within the packet.

     \item \textbf{Control Groups:} Control Group hooks allow programs to be logically tied to
a container context. Ingress and egress hooks allow filtering traffic. Egress
hook also allows limiting outgoing bandwidth by setting the Earliest Departure
Time~\cite{netdev} for the packet. Other hooks exist for control path operations.
Hooks for \cc{bind}, \cc{connect}, \cc{sendmsg}, \cc{getsockname},
\cc{getpeername} system calls allow enforcing policies and manipulating the
local or remote address passed in from user space. Hooks for \cc{setsockopt} and
\cc{getsockopt} allow changing and setting additional socket options based on
the desired policy~\cite{cgroups}.

     \item \textbf{User-Level Protocol:} The ULP hook (also known as SK\_MSG) allows invoking
\sys programs when a \cc{sendmsg} or \cc{sendfile} operation occurs for a
socket. These programs are used to enforce policies for the payload of each
message being sent~\cite{ciliumP}. Paired with kTLS~\cite{ktls}, it provides
transparent enforcement of ULP layer policies even for encrypted traffic
~\cite{borkmann2018combining}.

     \item \textbf{Congestion Control:} \sys programs can be registered as callbacks for TCP
congestion control operations ~\cite{Kernelop}, allowing faster experimentation,
data collection, and iteration of custom congestion control algorithms in
production.

     \item \textbf{Socket Operations:} These hooks allow attaching \sys programs to a TCP
socket's state transition events for example listen, connect, active connection vs
passive connection establishment, TCP header options~\cite{Cilium2024,Corbet2014}. It
allows using the peer address to dynamically select socket options and
congestion control settings, the retransmission timeout, maximum acknowledgment
delay timeout, and manipulating TCP header options.

\end{itemize}
\subsection{Profiling}
Applications use \sys programs attached to \cc{perf} events to collect data from
hardware performance counters and capture stack traces for the kernel and user
space. Due to its low overhead, \sys is used in multiple continuous profiling
applications~\cite{Aviv2023} without degrading workload performance significantly.

\subsection{Tracing}
Kernel functions and tracepoints can be traced by attaching \sys
programs which execute at function entry and exit. The programs have access to
all arguments of the kernel function or tracepoint. Low overhead instrumentation
(due to runtime code modification to emit direct calls to the program) makes
runtime tracing practical even for high-performance production workloads. A large set of tools use this facility to
perform runtime data collection, performance analysis, and profiling of kernel
subsystems \cite{Brendan2019,githubbcc}.

\subsection{Security}
The LSM BPF subsystem~\cite{LSMBPF, KRSI} allows \sys programs
to be attached to LSM hooks within the kernel. Programmable LSM hooks allow
enforcing security policies and auditing of the system. Object-local storage
maps~\cite{BPFMaps} are used to associate policy-specific data with a kernel
object (\eg~cgroups, inodes, sockets, etc.) acting as a subject of an LSM hook.
Additionally, LSM BPF programs can be attached to a cgroup context to constrain
the policy's scope to it~\cite{Stanislav2022}. These capabilities allow
LSM BPF to serve as a basic building block for flexibly building higher-level
security frameworks~\cite{010Findlay2020}. 

\subsection{Emerging}
Listed below are some emerging applications and innovations where \sys is leveraged to extend functionality beyond traditional use cases within the Linux kernel.

\begin{itemize}
    \item \textbf{Device Drivers:} The HID-BPF framework~\cite{HIDBPF} allows parts of the HID
device drivers to be implemented using \sys programs to filter events, make
driver fixes without changing the kernel, and inject additional input events.

     \item \textbf{Scheduling:} The ghOSt scheduler \cite{humphries2021ghost} attempts to
delegate scheduling to a user space application by using \sys to communicate
scheduler events over shared memory. SCHED-EXT~\cite{Theexten60, Tejun2022}
takes a different approach, by completely implementing the scheduling logic
within \sys programs as synchronous callbacks for the Linux scheduler.

     \item \textbf{Storage:} XRP~\cite{zhong2021bpf, zhong2022xrp} accelerates storage
applications by attaching \sys programs to the NVMe driver layer. The programs
then issue read operations directly, bypassing the kernel's storage stack, while
still keeping file system state in sync.
\end{itemize}
\section{Challenges}
\label{s:challenges}
In this section, we focus our attention on various challenges concerning \sys's
current design.

\subsection{Usability}
Navigating the intricacies of connecting \sys programs to attach points in the Linux kernel can be challenging. Understanding the various hook points available, discerning their suitability for specific tasks, and seamlessly integrating \sys programs with them necessitate a profound understanding of kernel intricacies and eBPF methodologies~\cite{ebpfdevchallenges}.

Furthermore, the dearth of exhaustive documentation and user-friendly development tools exacerbates these usability hurdles. Developers often find themselves grappling to locate pertinent resources and guidance, impeding their progress and efficiency in \sys development endeavors. Moreover, there are concerns about compatibility and stability across different versions of the kernel. Alterations in kernel APIs or hook placements can disrupt the behavior of \sys programs, compelling developers to continuously adjust their code for compatibility. This perpetual adjustment introduces unwelcome complexity and overhead~\cite{karim2023}.

\subsection{Scalability of the Verifier} A more scalable verifier directly translates
to increased capabilities for \sys, as a larger set of valid programs can pass
through it. The current verifier's static analysis theory is not capable enough
to handle extreme pessimistic cases of path explosion. Scalable handling of
loops is also critical for increasing expressiveness of programs. Instead of
unrolling, techniques such as loop invariant analysis~\cite{bpfbound2} or
summarization need further exploration.

\subsection{Correctness of the Verifier}
The safety guarantees of \sys hinge on the correctness and soundness of the
verifier's implementation.
The \sys verifier serving as the final arbiter deciding whether programs are
safe to execute comes with its own set of problems.  Firstly, the verifier
cannot be too conservative during its analysis, since that leads to
the rejection of a large set of valid and safe programs.  Secondly, the
verification algorithm should ideally terminate within a fixed amount of time,
without breaching the complexity limit during the program's symbolic
execution.

In practice, the verifier can minimize the impact of these vexing issues to a
great extent using the algorithms illustrated before in \autoref{sss:pruning}.
Any logical bug in the verification algorithm, failure to account for
unsafe program behavior in unexplored paths due to eager redundant path pruning,
or failure to capture and enforce kernel-specific invariants correctly directly
translate to unsafe or malicious \sys programs passing through the verifier,
undermining \sys's strong safety guarantees. The complexity and sheer size of
the verifier's codebase, paired with the high frequency of changes made every
kernel release to meet the needs of ever-increasing use cases make preserving
the correctness of the verification logic an increasingly daunting task for \sys
developers. 

\subsection{Formal Verification}
There has been no comprehensive formal investigation of the verifier and whether
its safety guarantees are sound. This remains an open research problem, and also
a huge undertaking due to large number of features supported by it. The
difficulty is further compounded by the high rate of changes made to the
verifier's codebase in every kernel release~\cite{bpf_verifier}.

Nevertheless, some promising attempts have been made thus far. Vishwanathan
\etal~\cite{tnumvish} formally specify the verifier's numerical abstract domain
(\autoref{sss:tnum}), providing soundness and optimality proofs. A part of this
work has since been adopted in the Linux kernel~\cite{tnumcommit}. Bhat~\etal~\cite{bhat2022formal} create an automated formal verification framework that
verifies the correctness of the C implementation of the range analysis logic
(\autoref{sss:tnum}) against a specification describing the correctness
invariants.  Nelson \etal~\cite{nelsonserval} create the Serval framework to
produce an automated verifier for the \sys instruction set. Nelson \etal
~\cite{nelsonjit} also apply automated proof techniques to verify \sys's JIT
implementations, by constructing a JIT correctness specification.

\subsection{Security}
The criticality of \sys bugs affecting the safety, integrity, and security of
the kernel, and by extension the rest of the system is self-evident from the set
of reported vulnerabilities in the past, which repeatedly abused it as a potent
vector for exploitation~\cite{Matan2022} in unprivileged mode. As a consequence, \sys
developers chose to disable the unprivileged mode by default~\cite{Disallowunpriv}, and
even when enabled explicitly, it is extremely restrictive to reduce the kernel's
attack surface exposed to untrusted and potentially
malicious users. Hence, \sys today remains largely useful only in the trusted
user model. 

The introduction of \texttt{CAP\_BPF} with linux 5.8 aims to separate BPF functionality from the broader \texttt{CAP\_SYS\_ADMIN} capability. The general idea is that a user who has this capability is able to (among other things) create bpf maps or load \texttt{SK\_REUSEPORT} programs. However, capabilities like \texttt{CAP\_NET\_ADMIN} and \texttt{CAP\_PERFMON} are still required for loading networking and tracing programs respectively,highlighting the ongoing need for elevated privileges in certain eBPF operations. 

Improving the correctness guarantees of the \sys verifier and JIT implementation
might, among other things, allow for the possibility of relaxing the restrictions
imposed on the \sys programs in the unprivileged mode, making \sys more useful
for a larger set of use cases which do not require privileges~\cite{seccompbpf}.

\subsection{Code Reuse}
Code reuse in eBPF programs is a double-edged sword, because on the one hand there is CO-RE, as explained in \autoref{par:co-re}, which allows to use the same compiled programs on different Linux versions by fixing memory offsets for data structures at load time.
On the other hand, although function calls are allowed in eBPF programs, there is no support for static or dynamic libraries. This means that two or more programs living in separate source files may need to reimplement the same functions.

\section{Conclusion}
\label{s:conclusion}

We examine the operational mechanisms of different components within the \sys subsystem in the Linux kernel, focusing on its objectives to offer safety assurances and analyzing the current implementations of these objectives. We discuss how various components in the \sys ecosystem come together to enable enhanced capabilities that users leverage to improve system performance, implement observability, security, and monitoring infrastructure, and enable
high performance networking applications.  We discuss challenges concerning
\sys's current direction and future progress, and the open research questions
presenting themselves.

\sys has seen wide adoption since its introduction into the Linux kernel.
It remains under active development. While its design is guided by a strong
focus on safety, flexibility, and performance, the primary driving force behind
\sys's continued evolution are its users within and outside the kernel.

More importantly, we illustrate how \sys empowers users to rethink conventional
operating system design. It allows users to perform radical changes to core
kernel subsystems with relative ease and confidence, and innovate quickly in a
fashion which was impractical before for a popular OS kernel like Linux.

\section*{Acknowledgements}
A special thanks to Kumar Kartikeya Dwivedi, who was instrumental in initiating the work on this paper and early in the writing process, but who unfortunately was not able to participate in the final phases of preparing this manuscript. This research was funded by Red Hat Research in the 
 \href{https://research.redhat.com/blog/research_project/security-and-safety-of-linux-systems-in-a-bpf-powered-hybrid-user-space-kernel-world/}{https://research.redhat.com/blog/research\_project/security-and-safety-of-linux-systems-in-a-bpf-powered-hybrid-user-space-kernel-world/} project.

{\footnotesize \bibliographystyle{acm}
\bibliography{bibliography}}
\appendix
\appendix
\section{eBPF Operations}
\label{s:appendix-ebpf_ops}
Table \ref{tab:ebps:operations} is a list of common \sys operations along with their descriptions~\cite{insnset}.

\begin{table}[ht!]
\footnotesize
\caption{List of eBPF Operations}
\centering
\begin{tabular}{|p{70pt}|p{150pt}|}
\hline
\textbf{Operation} & \textbf{Description} \\
\hline
BPF\_ALU\_ADD & Add two registers \\
BPF\_ALU\_SUB & Subtract two registers \\
BPF\_ALU\_MUL & Multiply two registers \\
BPF\_ALU\_DIV & Divide two registers \\
BPF\_ALU\_MOD & Modulus of two registers \\
BPF\_ALU\_OR & Bitwise OR of two registers \\
BPF\_ALU\_AND & Bitwise AND of two registers \\
BPF\_ALU\_LSH & Shift left of register \\
BPF\_ALU\_RSH & Shift right of register \\
BPF\_ALU\_NEG & Negate value of register \\
BPF\_ALU\_XOR & Bitwise XOR of two registers \\
BPF\_ALU\_MOV & Move a value from one register to another \\
BPF\_ALU\_ARSH & Arithmetic right shift of register \\
BPF\_ALU\_END & End marker for ALU operations \\
BPF\_JMP\_JEQ & Jump if equal \\
BPF\_JMP\_JNE & Jump if not equal \\
BPF\_JMP\_JA & Jump always \\
BPF\_JMP\_JGT & Jump if greater than \\
BPF\_JMP\_JGE & Jump if greater than or equal \\
BPF\_JMP\_JLT & Jump if less than \\
BPF\_JMP\_JLE & Jump if less than or equal \\
BPF\_JMP\_JSET & Jump if bitwise AND with immediate is true \\
BPF\_JMP\_CALL & Call function \\
BPF\_JMP\_EXIT & Terminate execution \\
BPF\_JMP\_ALU64 & Perform 64-bit arithmetic and jump \\
BPF\_JMP\_X & Reserved for future use \\
BPF\_JMP\_ADD & Add offset to register \\
BPF\_JMP\_MUL & Multiply register by scale \\
BPF\_JMP\_NEG & Negate register \\
BPF\_JMP\_AND & Bitwise AND with register \\
BPF\_JMP\_OR & Bitwise OR with register \\
BPF\_JMP\_XOR & Bitwise XOR with register \\
BPF\_JMP\_MOV & Move register to another \\
BPF\_JMP\_ARSH & Arithmetic right shift of register \\
BPF\_JMP\_END & End marker for JMP operations \\
BPF\_STX & Store into memory \\
BPF\_LDX & Load from memory \\
BPF\_ST & Store value \\
BPF\_LD & Load value \\
\hline
\end{tabular}
\label{tab:ebps:operations}
\end{table}

\section{Control-Flow Graph}
\begin{figure*}[]
	\centering
	\includegraphics[width=0.8\textwidth]{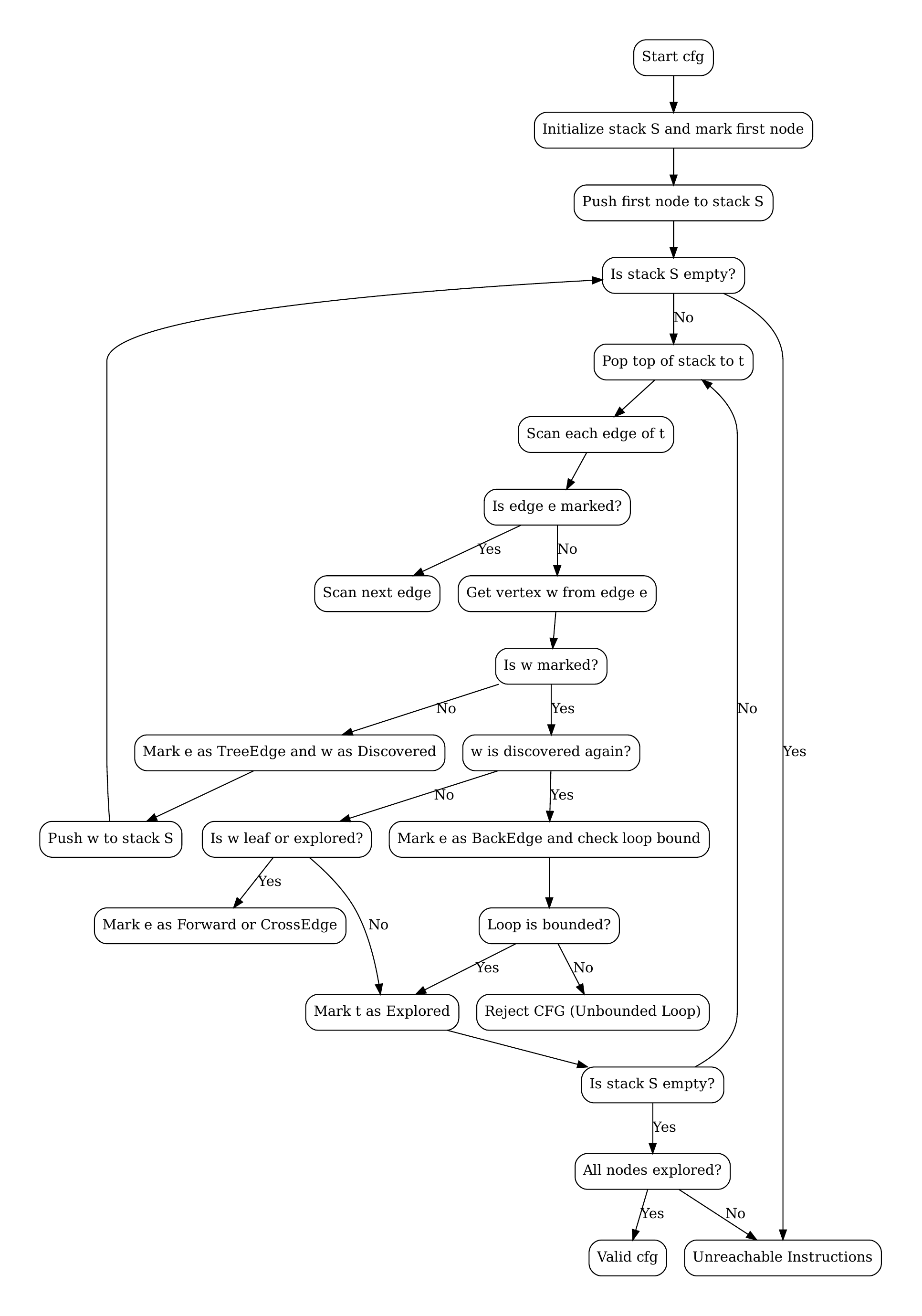}
	\caption{The flowchart illustrates the \texttt{check\_cfg()}\cite{bpf_verifier} function of the \sys verifier used to verify a Control-Flow Graph (CFG). It begins by initializing a stack (S) and marking the first node. Through depth-first search (DFS), it explores nodes (t) and scans edges (e). Edge classifications such as tree-edge, back-edge, and Forward/cross edge are determined based on traversal states. The function ensures all nodes are explored and checks for unreachable instructions.}
	\label{s:appendix-fig:cfg}
\end{figure*}
\end{document}